\documentclass[ASNA,twocolumn]{USG} 
\usepackage{anyfontsize} %
\usepackage{graphicx,color,epsf}
\usepackage{hyperref}
\usepackage{xcolor}
\usepackage{amsmath}
\usepackage{amssymb}
\usepackage{colortbl}   
\usepackage{xcolor}     

\usepackage{steinmetz}
\graphicspath{{./images/}}
\specialissue{}

\articletype{RESEARCH ARTICLE}%
\subarticletype{}

\received{XX Month Year}
\revised{XX Month Year}
\accepted{XX Month Year}
\journal{Macromolecular Theory  and Simulations}
\volume{0}
\copyyear{Year}
\startpage{1}
\articledoi{10.1002/0000}

\begin{document}
\title{Perimeter shell properties of branched polymer adsorbents in various { conformational states}}

\author[1]{Viktoria Blavatska}
\author[2,3]{Jaroslav Ilnytskyi}

\authormark{BLAVATSKA \textsc{et al.}}
\titlemark{PERIMETER SHELL PROPERTIES OF BRANCHED POLYMER ADSORBENTS WITH VARIOUS LEVEL OF FLEXIBILITY}

\address[1]{\orgdiv{Laboratory for Statistical Physics of Complex Systems
		, }\orgname{Yukhnovskii Institute for Condensed
		Matter Physics of the National Academy of Sciences of Ukraine, }%
\orgaddress{\state{Lviv, }\country{Ukraine}}}
\address[2]{\orgdiv{Department for Computer Simulations of Many-Particle Systems
		, }\orgname{Yukhnovskii Institute for Condensed
		Matter Physics of the National Academy of Sciences of Ukraine, }%
\orgaddress{\state{Lviv, }\country{Ukraine}}}
\address[3]{\orgdiv{Institute of Applied Mathematics and Fundamental Sciences, }\orgname{Lviv Polytechnic National University, }%
\orgaddress{\state{Lviv, }\country{Ukraine}}}

\corres{V. Blavatska  (\email{viktoria@icmp.lviv.ua})}


\presentaddress{}

\fundingInfo{
	 National
	Research Foundation of Ukraine,   Grant/Award Number: 2025.07/0366.}

\keywords{Polymers | Lattice models | Conformational properties | Monte Carlo simulations | Adsorption}
\transkeywords{Polymers | Lattice models | Conformational properties | Monte Carlo simulations | Adsorbtion}

\abstract[ABSTRACT]{The structural accessibility and perimeter shell properties of branched macromolecules
	are studied using Monte Carlo simulations on a simple cubic lattice. {  Constituent polymer chains of these  macromolecules are constructed in an algorithmic way as: a set of collinear monomers; a self-avoiding walk (SAW); tigthly packed monomers. These mimic the ground states for a rigid rod, flexible chain, and collapsed chain conformational states, respectively.}  We analyzed the set of observables  that are related to adsorption, such as perimeter shell size (number of available sites for particle adsorption) per monomer $\langle \bar{n}_a \rangle$,   { mean coordination number of perimeter shell} $\langle n_{{\rm bond}} \rangle$ (average number of contacts per adsorbed particle), and coordination accessibility per monomer $\langle \bar{n}_s \rangle$.
	Our results demonstrate that branching reduces the perimeter shell size while it increases  { mean coordination number of perimeter shell} across all conformational states, due to higher local monomer density near branching cores and formation of shared adsorption sites or microcages. Stiff extended structures yield the highest values for $\langle \bar{n}_a \rangle$ (approaching $4$ in the limit of long chains), whereas the flexible ones provide the largest values for $\langle n_{{\rm bond}} \rangle$. We can note that the { collinearity of monomers in constituent chains } works in the same direction as reduction of branching ($\langle \bar{n}_a \rangle$ grows), whereas the { tight packing of them} -- in the same direction as the increase of branching ($\langle \bar{n}_a \rangle$ decreases).}

 \transabstract[transABSTRACT]{.}

\abbr{SAW, self-avoiding random walk; PERM, pruned-enriched Rosenbluth method.}

\contributed{V. Blavatska and Ja. Ilnytskyi contributed equally to this study.}

\copyright{Copyright Statement
\\[5pt]
  ©  2024 XXXX}


\maketitle


\section{\label{sec1}Introduction}

Selective adsorption of organic and inorganic pollutants from air, water and soil, aimed at their cleaning and/or purification, is of great practical importance \cite{Aigbe2023, Czech2025}. Despite the differences between the chemical and physical mechanisms in each particular case, the adsorption effectiveness dictates rather similar requirements to the adsorbent: mechanical stability, maximum surface area, specific arrangement and accessibility of adsorbing chemical groups, etc. These requirements can be addressed more readily for the polymer-based adsorbents \cite{Pan2009, Alkhaldi2024},  where  the polymer macromolecules serve as stationary adsorbent medium (matrix) for external targets. The examples include such superadsorbents as macroporous resins, biopolymer-based adsorbents, hydrogels, all representing the same class of highly crosslinked polymer networks \cite{Martn2011, MasoumiSahar2024, Cheng2024}. To achieve the desired structure of a polymer adsorbent, and, consequently, its high effectiveness, one can vary the molecular mass and flexibility of its constituent chains, the amount and type of molecular branching, and other details. Clarification of the interdependence between all these factors could lead to prediction of the optimal molecular structure of a polymeric adsorbent.

Comparison between linear and branched polymer chelating absorbers for removal of heavy metal ions and hazardous dyes have been done in a range of experiments \cite{Kobayashi1987, Yi2024, Blanco2009, Liu2017, Jo2022}. Branching was found to enhance the effectiveness of chelation because of two factors: natural formation of the chelation-adsorption-friendly cages around the branching points, and spatial separation of individual linear spacer-chains which increases their accessibility for pollutants \cite{Mouhamad2023}.

Typical polymer networks and gels are crosslinked randomly and, therefore, are characterized by a polydispersity in both branching points multiplicity and polymer spacer lengths. The effects of the branching points of a particular kind, and of the spacers of a given length, are all mixed up and cannot be separated easily. On the other hand, the macromolecules of a regular branching pattern such as: star-like, pom-pom, snowflake, dendrimer, etc. \cite{Gao2004, Xu2015, Jeon2018, Seo2020}, allow such a possibility. Therefore, such molecular architectures are logical candidates for studying the effects of various types of branching on the effectiveness of adsorption.

\begin{figure}[!h]
\begin{center}
\includegraphics[width=52mm]{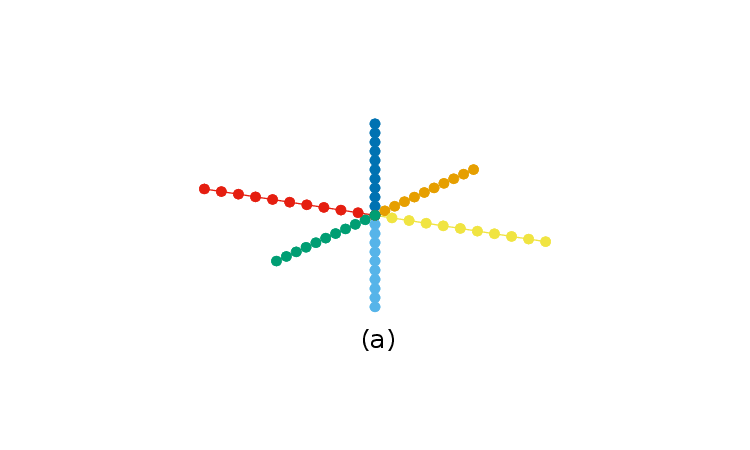}
\hspace*{-2.5cm}
\includegraphics[width=52mm]{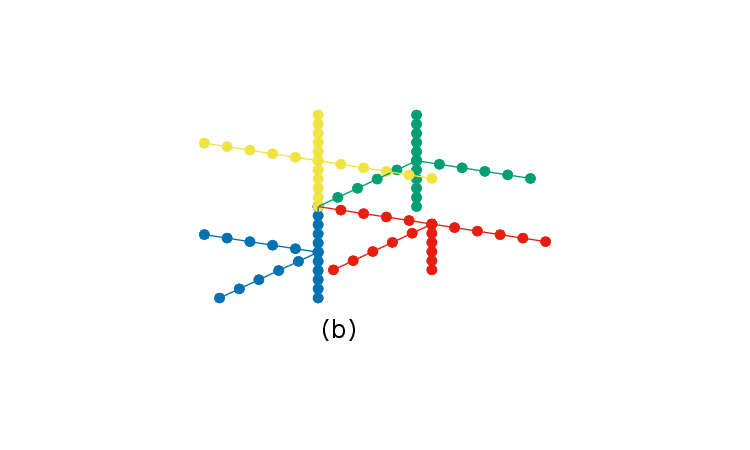}\\
\includegraphics[width=52mm]{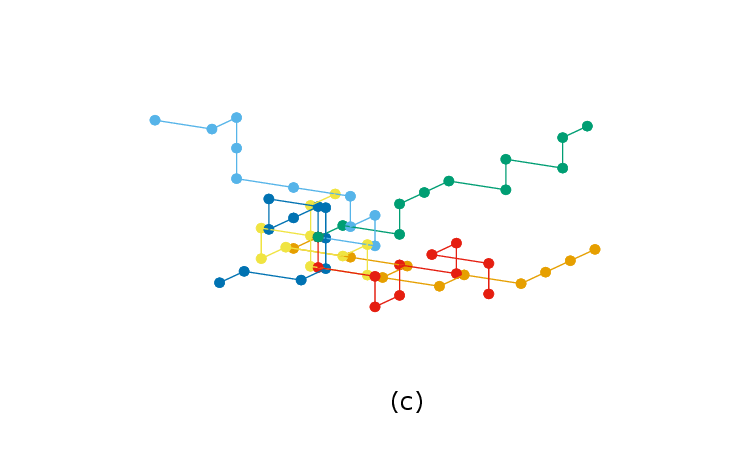}
\hspace*{-2.2cm}
\includegraphics[width=52mm]{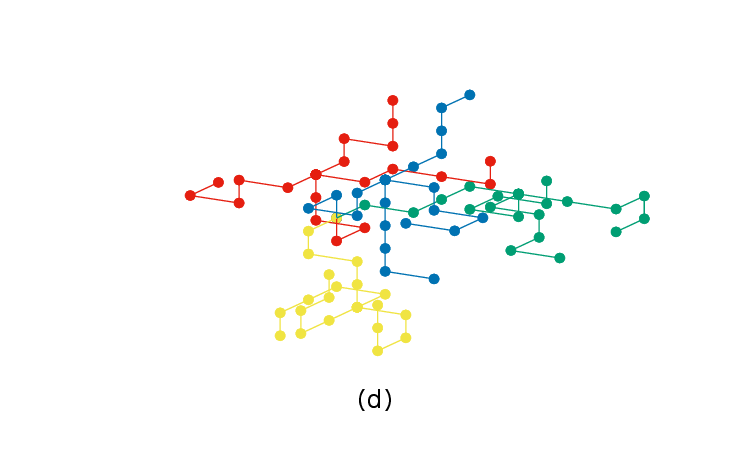}\\
\includegraphics[width=45mm]{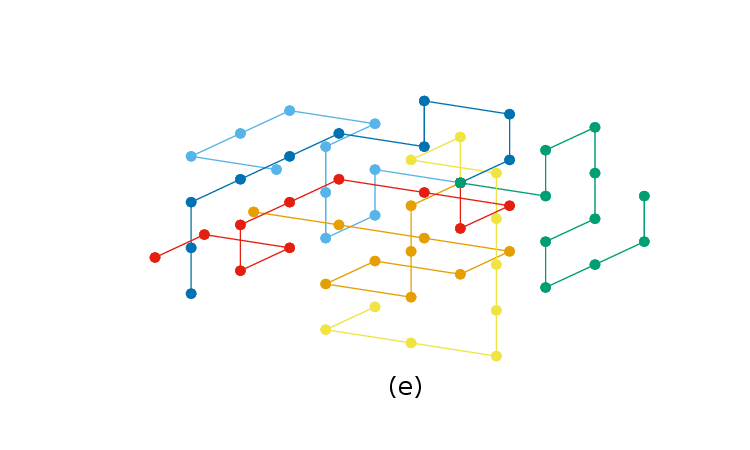}
\hspace*{-1.0cm}
\includegraphics[width=45mm]{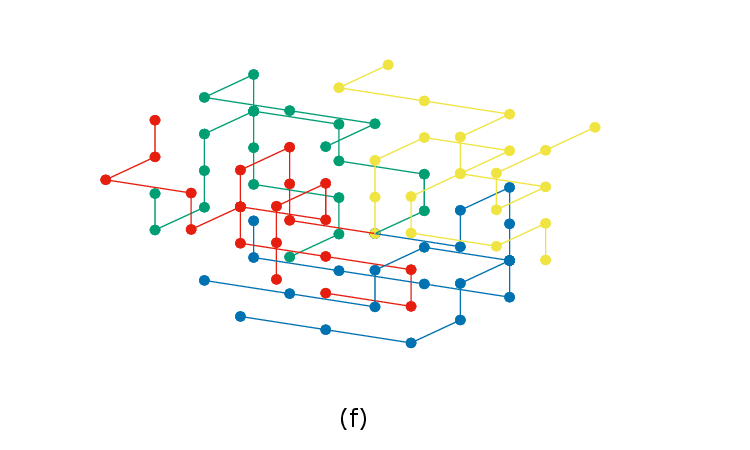}
\end{center}
\caption{\label{starscheme}{Monte Carlo simulation snapshots demonstrating the spatial conformations of representative polymer structures on a simple cubic lattice. (a), (c), and (e):   a star polymer with $f=6$ branches;   (b), (d), and (f):  a snowflake polymer with $f=4$ branches and $g=3$ twigs. (a) and (b): polymers comprising rigid rod branches; (c) and (d): flexible polymers; (e) and (f): tightly packed polymers.}}
\end{figure}

Star polymer represents the simplest example of a branching molecular architecture with, typically, a single branching point \cite{Zimm1949, Likos2001}. It contains    $f$ linear branches (the star functionality) that radiate outwards from a central core (Fig. \ref{starscheme}a), thus reconstituting a single linear chain at $f=1$. Such polymer exhibits a more closely packed object than its linear counterpart of the same molecular weight \cite{Miyake1983, Wei1997, vonFerber1997, CM, Kalyuzhnyi2018}. 
A snowflake polymer topology \cite{Liu2021, Leitner22, Haydukivska2023a}, that can be interpreted as a first-generation dendrimer, is of particular interest. It consists of a  star polymer of the functionality $f$ with additional $g$ twigs radiating from the terminal points of each arm (Fig. \ref{starscheme}b).

Besides the type of branching of a particular polymer, there is another factor that affects its structure: the conformation state of its constituent chains,
reflected in the scaling law for their mean-square radius of gyration $\langle R_g^2 \rangle$. In general,
\begin{eqnarray}
	\langle R_g^2 \rangle &\sim& N^{2\nu},\label{nur} 
\end{eqnarray}
where $N$ is the number of monomers in a chain, and $\nu$ is the critical index for a correlation length. $R_g$ can be measured directly via static light scattering \cite{Ferri2000, Smilgies2015}, and distinguishes between {  different conformational states of constituent polymer chains}. Additional characteristic is the asymmetry measure of typical polymer configurations, such as the average asphericity $\langle A \rangle$ \cite{Aronovitz1986, Rudnick1986}. This quantity is bound between a maximum value of one for a completely rigid, rod-like conformation, and a minimum value of zero for an ideally spherical shape, thereby obeying the inequality    $0 \leq \langle A \rangle \leq 1$. Hereafter, $\langle \ldots \rangle$ denotes an ensemble average over all possible polymer conformations.

The SAW mimics a polymer chain in a good solvent, where it  adopts the shape of a coil \cite{deGennes1979, desCloiseaux1991}, and, assuming that the adjacent monomers are separated by at least the Kuhn segment length, the scaling index is $\nu\approx 0.588$ in $D=3$ ($D$ being the space dimensionality) \cite{Guida1998}. We note that this law holds also for the macromolecule, composed of flexible chains, as a whole.
The asphericity of a macromolecule $A$, however, is influenced strongly by the molecular topology. For instance, for the case of the $f$-functional star, the limit case of a linear chain exhibits significant spatial anisotropy with $\langle A(f{=}1) \rangle = 0.431(2)$ \cite{Jagodzinski1992}, whereas the star with $f>1$ is characterized by a gradual decrease of its asphericity with the increase of its functionality, $f$:
$A(f{=}3)=0.306(1)$, $A(f{=}4)=0.227(1)$, $A(f{=}5)=0.177(1)$, $A(f{=}6)=0.140$  \cite{Bishop1991}.

{ Tightly packed polymer chains can be associated with the case when they are surrounded by a poor solvent}, in which case they collapse into a dense globular structure. According to the polymer physics and the scaling theory, it can have two different forms \cite{Yamakawa1971, Binder2008}: a disordered globule with high uniform concentration, and a highly symmetric crystalline or semi-crystalline form (e.g. lamellae).  
This dense compact state is driven by a short-range, solvent-mediated monomer-monomer attraction, whereas conformation entropy works against the polymer collapse.
{  From structural point of view, one may say that such polymers tend to form the mostly compact, tightly packed structure with maximum number of monomer-monomer contacts, which implies possesing a high level of bending.}
The effective size measure (\ref{nur}) of a polymer in this state scales with $\nu=1/D$.  The shape of linear and star-branched polymers in compact state have been questioned by means of the Monte Carlo simulations  in Refs.~\cite{Bishop1989,Sikorski1998}, indicating a substantial decrease in the asphericity parameter    $\langle A \rangle$, driven by the higher symmetry of compact globules as compared to crumpled coils in good solvent regimes.

{ The collinear polymer chain represent a semiflexible polymer in a limit case of infinite rigidity.}  The latter was  introduced by Flory \cite{Flory1956} in a study of polymer crystallization. In this case, bending of two adjacent bonds causes the energy penalty, quantified via a stiffness parameter. Depending on solvent conditions, semiflexible polymers can adopt a variety of ordered morphologies, including rod-like or racquet-like conformations, as the result of the competition between the monomers attraction and chain stiffness \cite{Montesi2004, Bloomfield1991, Araki2016, Amiri2026}. In particular, the polyelectrolytes are more rigid because of long-range electrostatic interactions between the monomers, and these tend to behave like a semi-flexible rods in good or theta-solvents \cite{ Araki2016}. The  rigid conformations of polymers with exceptional stiffness are often found in aromatic or heterocyclic backbones \cite{Hu2003, Guan2022, Abdous2026}. In this regime, the radius of gyration  scales linearly with  $N$, so that  $\nu=1$.  Rigid-rod polymers are of interest because of their high mechanical and thermal/oxidative stability, as well as environmental resilience.  Experimental studies of semiflexible and rigid star polymers reveal their unique features in molecular recognition and self-assembly behavior \cite{Miyabe2011, Kikuchi2016, Hasegawa2017}. The rigid branched dendrimeric structures were synthesized in Ref.~\cite{Freudenberger1994} based on poly([1.1.1]propellane) and  in Ref.~\cite{Konkolewicz2010}, based on fibrin polymerization.

In the present study, we focus on the perimeter shell properties of branched macromolecules that may be related to their ability to adsorb small particles.
Their constituent chains are constructed in an algorithmic way as: a set of collinear monomers; a self-avoiding walk (SAW); tigthly packed monomers. These mimic the ground states for the limit cases of a rigid rod, flexible chain, and collapsed chain conformational states, respectively. 
Our analysis is performed using the discrete model of a growing polymer chain on regular lattices within the  framework  of self-avoiding random walk - SAW \cite{Kremer1988,Binder2012, Hooper2020} which captures the essential structural properties of macromolecules on a macroscopic scale without requiring a detailed description of intermolecular potential. 
Within the framework of this model, each consecutive step of SAW trajectory represents a single monomer of the polymer structure.
This approach has been successfully applied in our previous studies \cite{Blavatska2024a,Blavatska2024b} focusing on the properties that can be related to adsorption processes involving linear and branched polymer architectures.
Here, we focus on the perimeter of SAW, as given by the set of sites occupied by SAW trajectory with at least one vacant lattice site in their neighborhood. 
The perimeter shell is thus defined as the set of vacant lattice sites adjacent to the perimeter sites; related to the concept of  ``atmosphere" of lattice clusters \cite{Rechnitzer2008,JansevanRensburg2008,Owczarek2008,Alvarez2009}. 
We focus on the following structural characteristics of perimeter shell:

\begin{itemize}
\item  perimeter shell size per monomer  $\bar{n}_{a}$: The average number of vacant sites that are adjacent to at least one monomer of the perimeter, normalized by the total number of monomers $N$.

\item  coordination accessibility per monomer
 ${\bar{n}}_{s}$: The average total number of open neighboring directions for all the monomers of the perimeter normalized by $N$;

\item   { mean coordination number of perimeter shell} $n_{\mathrm{bond}}$: 
The average number of polymer monomers adjacent to a single perimeter shell site. 

\end{itemize}

By interpreting the perimeter shell as an adsorbent matrix, these parameters are related to experimentally observable adsorption characteristics as follows.
$\bar{n}_{a}$ represents the theoretical maximum number of particles that can be potentially adsorbed at the  neighborhood of a single monomer and can be associated with  maximum adsorption capacity; 
${\bar{n}}_{s}$ gives the maximum number of potential coordination bonds between a single monomer and adsorbed particles; $n_{\mathrm{bond}}$ corresponds to the averaged number of bonds established  between a single adsorbed particle  and the adjacent  monomers of a perimeter, related in the same way to the concept of adsorption coordination strength { (an analogue of binding energy in the case of discrete lattice or off-lattice models with explicit pairwise attractive potential between monomers and particles)}.

 These parameters can be mapped onto realistic disordered environment where the adsorbent polymer matrix captures target particles (pollutants) from a surrounding solution. Within the formalism of a lattice model, we can introduce point-like defects with a concentration    $c$ ($0 \le c \le 1$). 
  At smaller $c$,  the  obstacles fill the available sites randomly with probability proportional to $c$. 
Thus,  the average number of actual adsorbed particles per monomer  $n_a(c)$ is assumed to scale as $n_a(c)\sim c\bar{n}_{a}$, while the average number $n_s(c)$ of active bonds formed between the monomer and the adsorbed particles:   $n_s(c)\sim c  \bar{n}_{s}$.
{  It is important to mention that assumed linear scaling  with the defect concentration $c$ is a phenomenologically observed in the dilute low-density limit of a non-interacting adsorbate e.g. in our previous study \cite{Blavatska2024a}. Our lattice parameters $\overline{n}_{a}$  and $\overline{n}_{s}$  represent strictly the geometric upper bounds (the total number of available sites and coordination directions in a site-filling scenario), rather than describing the entire functional range of a thermodynamic adsorption isotherm.}
 
Note that the mean coordination number   $n_{\mathrm{bond}}$ is an intrinsic geometric property of the polymer's local conformation and remains invariant with respect to the defect concentration    $c$.

The remainder of this paper is organized as follows. Section \ref{II} introduces the lattice model of the growing polymer chain and details of the calculation of observables. Section \ref{III} provides a thorough analysis of the data obtained for branched polymer structures in rigid, fully flexible, and tightly packed regimes. Finally, Section \ref{IV} presents our conclusions and outlook.

\section{The method}\label{II}

To apply the polymer growth algorithm strengthened by pruned-enriched population control (PERM) \cite{Rosenbluth55,Grassberger97} for constructing the structures shown in Fig. \ref{starscheme}, we consider a three-dimensional ($3D$) simple cubic lattice. The first monomer of a polymer structure is chosen at the point with coordinates    $\vec{R}_0$. Next, a branched structure is generated by growing a predefined number    $f$ of the first generation chains (branches) that radiate from this core. 
The first monomer ($n=1$) of each linear branch is placed at a randomly selected unoccupied nearest-neighbor site of the central core    $\{x_0, y_0, z_0\}$, which is shared by all    $f$ branches. Consequently, for a simple cubic lattice, the number of branches is geometrically constrained to    $1 \leq f \leq 6$.

Each    $k$-th branch ($k=1,\ldots,f$) grows sequentially: its    $n$-th monomer ($n=1,\ldots,N$) is placed at a site with coordinates    $\vec{R}_n^k=(x_{n}^{k}, y_{n}^{k}, z_{n}^{k})$ chosen from the unoccupied nearest neighbors of the  $(n-1)$-th monomer (out of up to 5 available sites). 
To construct a snowflake configuration with    $g$ twigs, each terminal monomer of the primary branches ($\vec{R}_N^k$ with $k=1,\ldots,N$) is treated as a new starting core for $l=1,\ldots,g$ growing linear chains, such that $\vec{R}_N^{k}=\vec{R}_0^{k,l}$. The generation process then proceeds for each twig reaches its specified length $N$ with the monomer coordinates denoted as $\vec{R}_n^{k,l}=(x_{n}^{k,l}, y_{n}^{k,l}, z_{n}^{k,l})$. 
{   At this point it is important to note that our approach focuses on the three distinct conformational regimes (states) of polymer chains, generated exclusively by using qualitatively distinct algorithmic growth rules as listed below: }

\begin{itemize} 

\item  rigid polymer structures: 
Each branch is straight and runs along one of a randomly chosen spatial axis (chain bending is prohibited).

\item flexible polymer structures:  
The $n$-th monomer is any unoccupied nearest neighbor of the $n-1$-th one, resulting in the SAW construction algorithm for each chain.

\item tightly packed polymer structures:
     For each available neighbor site $i$ of $(n-1)$th monomer, its number of contacts  $z_i$ is calculated.  The site with the highest $z_{i}$  is selected for  $n$-th monomer. In the case when two or more sites share the same maximum number of contacts, one of them is chosen at random with equal probability.

\end{itemize}

\begin{figure}[!h]
\begin{center}
\includegraphics[width=52mm]{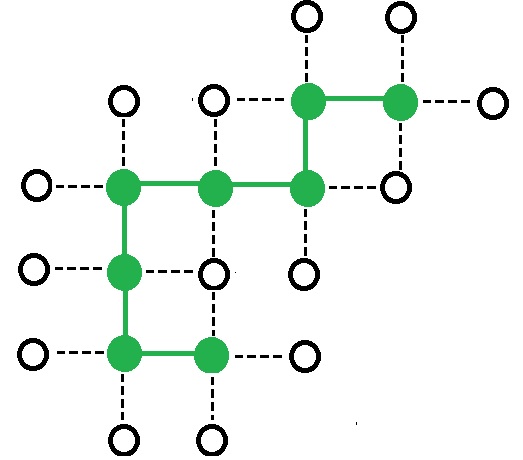}
\end{center}
\caption{\label{relation}  { Schematic representation of the  2D lattice polymer chain with $N=8$ monomers, shown by green filled circles. The vacant lattice sites adjacent to the monomers, shown with $N_a=14$ open circles,  constitute the perimeter shell. Dashed lines denote the available coordination directions that determine the total coordination accessibility ($N_s=18$). The resulting normalized metrics thus are: the perimeter shell size per monomer $\overline{n}_a = 14/8$,  the mean total coordination accessibility per monomer $\overline{n}_s = 18/8$, the mean coordination number per shell site  $\langle n_{{\rm bond}} \rangle = 18/14$.}
}
\end{figure}

A weight $W_n$ is assigned to each trial configuration at the $n$th step, given by the Rosenbluth weights \cite{Rosenbluth55}
\begin{equation}
W_n^{{\rm Rosenbluth}}= \prod_{l{=}1}^n m_l. \label{weight}
\end{equation}
Here, $m_l$ 
{  counts the number of unoccupied nearest neighbor sites in the restricted subset of sites that satisfy the corresponding algorithmic ``filters'' at step $l$. Namely, $m_{l}$ is equal to: (i) the number of  non-occupied sites along the chosen spatial axis in the rigid polymer regime;  (ii) all unoccupied nearest neighbors in the flexible polymer regime; and (iii) only the subset of unoccupied sites that corresponds to  the degenerate maximum number of contacts $z_{i}$  in the tightly packed regime.

Since the algorithm chooses one site uniformly at random from the $m_{l}$  allowed options at each step $l$, the conditional probability of making this choice is $1/m_l$. The total probability of generating the configuration at $n$th step is the product over all sequential growth steps:
\begin{equation}
P_n=\prod _{l=1}^{n}\frac{1}{m_{l}}.
\end{equation}
Thus, the Rosenbluth weight defined in Eq. (\ref{weight}) satisfies the condition $W_n = 1/P_n$.
}
The growth process continues until the required total number of monomers in each branch,  $N$, is reached. 
 The total number of monomers in constructed branched structures are thus given by $N_{{\rm total}}=1+fgN$ and $N_{{\rm total}}=1+(f+fg)N$ for star-like and snowflake topologies, respectively.

{ According to PERM population control, configurations with too small weights are pruned, while the sample is being enriching with copies of the high-weight configurations \cite{Grassberger97}. The thresholds values $W_n^{<}$ and $W_n^{>}$ are chosen; if the current weight $W_n$ of an $n$-monomer chain is less than $W_n^{<}$,  the chain is either discarded with probability $1/2$, or it is kept and its weight is doubled.  If $W_n$ exceeds  $W_n^{>}$, the configuration is doubled and the weight of each copy is taken as half the original weight. The pruning-enrichment control parameters $W_n^{<}$ and $W_n^{>}$  are adjusted in such a way that on average 10 chains of total length $N_{{\rm total}}$ are generated per each iteration. 
}

The size and the shape characteristics of a polymer are obtained conveniently from its gyration tensor, $\bf{Q}$, with the components defined as
\begin{equation}
Q_{ab}=\frac{1}{2N^2_{{\rm total}}}\sum_{n,m=1}^N\sum_{k,r=1}^{f}\sum_{l,s=1}^g (a_{n}^{k,l}-a_{m}^{r,s})(b_{n}^{k,l}-b_{m}^{r,s}),
\end{equation}
where $\{a,b\}=\{x,y,z\}$.

The squared radius of gyration $R_g^2$ is thus given by the trace of  $\bf{Q}$:
\begin{equation}
	R_g^2=\frac{1}{2N^2_{{\rm total}}}\sum_{n=1}^N\sum_{k,r=1}^{f}\sum_{l,s=1}^g (a_{n}^{k,l}-a_{m}^{r,s})^2. \label{rfdef}
\end{equation}
The asphericity parameter $A$ is evaluated based on the eigenvalues $\lambda_i$ ($i=1,2,3$) of the tensor $\bf{Q}$ according to \cite{Aronovitz1986,Rudnick1986}:
\begin{equation}
	A_d=\frac{1}{d(d-1)}\sum_{i=1}^d \frac{(\lambda_i-{\overline{\lambda}})^2}{ {\overline{\lambda}^2}}, \label{asfer2}
\end{equation}
with ${\overline{\lambda}}\equiv {\rm Tr}\, {\bf{Q}}/d$ being
the mean arithmetic eigenvalue of  the gyration tensor eigenvalues. 

Once a configuration with the specified parameters, $N_{{\rm total}}$, $f$, and $g$, is constructed, we assign a label  $O(x(j), y(j), z(j))$ to each $j$th lattice site  with coordinates $\{x(j),y(j),z(j)\}$ such that 
$$
O(x(j), y(j), z(j)){=}\begin{cases}
1, \, \,   \text{if $j$th site is occupied by monomer,}\\
0, \, \,   \text{otherwise.}
\end{cases} 
$$
 Then, the accessibility (i.e., the total number of potential adsorption sites) of the polymer structure is given by 
\begin{eqnarray}
&&N_s(N_{{\rm total}})=\sum_{j}\delta_{O(x(j), y(j), z(j)),\,1} \Big[\delta_{O(x(j)\pm1, y(j), z(j)),\,0}+\nonumber\\
&&+\delta_{O(x(j), y(j)\pm1, z(j)),\,0}+\delta_{O(x(j), y(j), z(j)\pm1),\,0}\Big]. \label{Nsdef}
\end{eqnarray}
so that the normalized value $\bar{n}_s=N_s(N_{{\rm total}})/N_{{\rm total}}$ represents the average
 mean total coordination accessibility 
(the potential number of outgoing active bonds) per single monomer. 

To estimate the total number $N_a(N_{{\rm total}})$ of unoccupied sites within the perimeter shell of a polymer structure with $N_{{\rm total}}$ monomers, we assign another label, $v(x(j), y(j), z(j))$ to each    $j$-th lattice site, initializing its value to zero. Then, the following condition is checked for each    $j$-th site:
 
 if $O(x(j), y(j), z(j))=1$, then
\begin{itemize}
    \item if $ O(x_j\pm1,y_j,z_j)=0$, then  
$v(x_j\pm1,y_j,z_j)=v(x_j\pm1,y_j,z_j)+1$;

\noindent
if $v(x_j\pm1,y_j,z_j)=1$ then $N_a(N_{{\rm total}})=N_a(N_{{\rm total}})+1$

 \item if $ O(x_j,y_j\pm1,z_j)=0$, then 
$v(x_j,y_j\pm1,z_j)=v(x_j,y_j\pm1,z_j)+1$;

\noindent
if $v(x_j\pm1,y_j,z_j)=1$ then $N_a(N_{{\rm total}})=N_a(N_{{\rm total}})+1$

\item if $ O(x_j,y_j,z_j\pm1)=0$, then 
$v(x_j,y_j,z_j\pm1)=v(x_j,y_j,z_j\pm1)+1$;

\noindent if $v(x_j,y_j,z_j\pm1)=1$ then $N_a(N_{{\rm total}})=N_a(N_{{\rm total}})+1$

\end{itemize} 
so that at the end of the procedure, $N_a(N_{{\rm total}})$ counts the total number of vacant sites within a perimeter shell of the polymer structure. The normalized value, $\bar{n}_a=N_a(N_{{\rm total}})/N_{{\rm total}}$, representing the perimeter shell size (that can be related to the maximum adsorption capacity) per single monomer, is introduced in a similar way.

Finally, the values of $v(x(i),y(i),z(i))$ give the  { mean coordination number of perimeter shell} $n_{{\rm bond}}$  (that can be interpreted as the adsorption coordination strength), the number of bonds that can be established between a particle located at site $i$  and the monomers of the polymer structure.

The configurational averaging for any observable of interest $O$ is given by
\begin{equation}
 \overline {O } =\frac{{\sum_{{\rm conf}} W_N^{{\rm conf}} O }} {Z_N} \label{confaver}
 \end{equation}
 with  $W_N^{{\rm conf}}$ being the weight of an $N$-monomer chain in a given configuration given by (\ref{weight}) and  ${Z_N} =\sum_{{\rm conf}} W_N^{{\rm conf}}$ is the partition sum.

{ 
To provide a clear geometric view on the introduced parameters, a schematic example of a polymer chain of length $N=8$ on a square lattice (for easier visualization) is given in Fig. \ref{relation}. The perimeter shell is formed exclusively by the adjacent unoccupied lattice sites (shown with open circles), yielding a total shell size of $N_a = 14$. The outgoing dashed lines illustrate the open, unoccupied neighboring directions from each monomer, contributing to the total coordination accessibility $N_s = 18$. To calculate the mean coordination number \(\langle n_{bond} \rangle\), we notice that  out of the 14 shell sites, 11 sites are adjacent to exactly one polymer monomer (single contact), 2 sites share contacts with two monomers (double contacts), and 1 mostly ``trapped'' perimeter site is enclosed by three monomers (triple contact). Summing these structural interactions gives \(\langle n_{bond} \rangle = (11 \times 1 + 2 \times 2 + 1 \times 3)/14 = 18/14\).

}

One easily notes that  the coordination accessibility $ \bar{n}_s$, perimeter shell size $  \bar{n}_a$, and the  { mean coordination number of perimeter shell}, $n_{\rm bond}  $, are related by the identity $ \bar{n}_s = \bar{n}_a\,\cdot  n_{\rm bond} $  \cite{Blavatska2024a}, thus these quantities are not fully independent. Therefore, in the subsequent discussions and presentation of results of our analysis, we primarily concentrate on the two key independent observables, $  \bar{n}_a  $ and $  {n}_{\rm bond}  $, which together provide a comprehensive description of the adsorption performance.

\section{Results}\label{III}

\vspace{0.5em}
{\bf A. Rigid polymers of various architecture}.\vspace{0.5em}\\
We begin by examining the observables of interest for the case of the polymers with the rod-like constituent chains. {These can be found among 
polyelectrolytes \cite{Araki2016}, aromatic or heterocyclic backbones \cite{Hu2003, Guan2022, Abdous2026}, polysaccharides \cite{Benalaya2024, Salehi2025}, etc. This case allows for simple analytical treatment. For a rod-like linear polymer chain of length    $N_{\text{total}}$, there are    $2$ end monomers with    $5$ available connections each, and    $N_{\text{total}}-2$ inner monomers with    $4$ connections each. Consequently, the mean total coordination accessibility per monomer  for a single monomer is given by:
\begin{equation}
\bar{n}_{s,f=1}^{{\rm stiff}}(N_{{\rm total}})=\frac{5\cdot 2 + 4(N_{{\rm total}}-2)}{N_{{\rm total}}}=\frac{2+4N_{{\rm total}}}{N_{{\rm total}}}. \label{stiffchainna}
\end{equation}
Following the same reasoning, we have $\bar{n}_a^{{\rm stiff}}(N_{{\rm total}})=\bar{n}_s^{{\rm stiff}}(N_{{\rm total}})$
and $n_{{\rm bond}}^{{\rm stiff}}(N_{{\rm total}})=1$.

 \begin{figure}[t!]
 \begin{center}
\includegraphics[width=8cm]{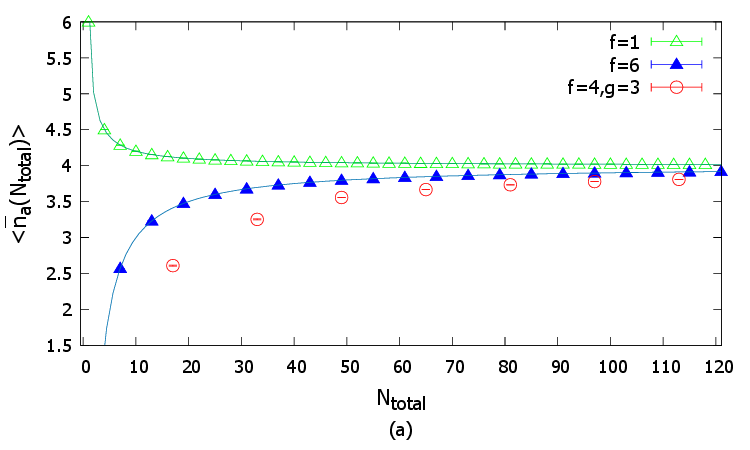}\\
\includegraphics[width=8cm]{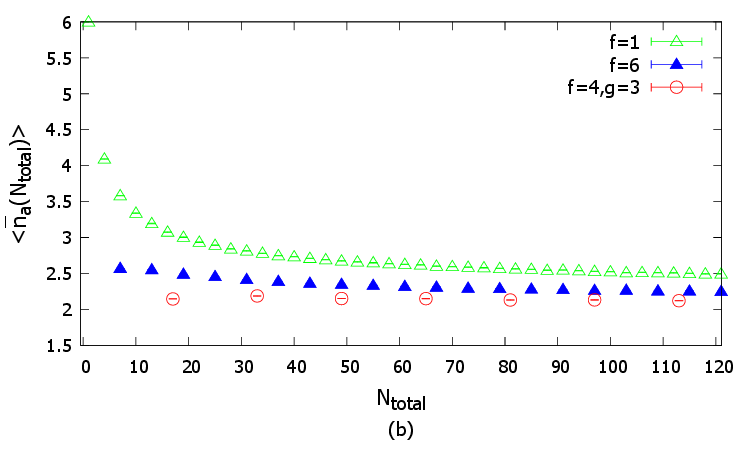}\\
\includegraphics[width=8cm]{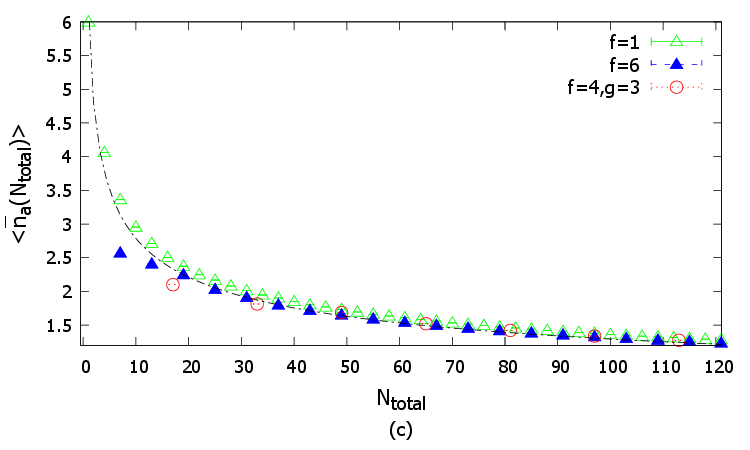}
\end{center}
\caption {\label{nan_all}  Perimeter shell size, $\langle \bar{n}_{a} \rangle$, for three representative polymer structures: comprising rigid rod chains (a), flexible (b), and tightly packed (c) regimes. Solid lines in (a) represent analytical evaluation as  given by Eqs. (\ref{stiffchainna}), (\ref{stiffstarna}), whereas in (c) the solid line gives a result of rough approximation of Eq. (\ref{nsglob}).
}
\end{figure}

For a star-shaped structure of the same total molecular weight,   $N_{\text{total}}$, consisting of    $f$ rod-like branches each of length    $N = (N_{\text{total}} - 1) / f$, we introduce the sequential numbering for the monomers along each    $k$-th branch (   $k = 1, \dots, f$) as    $n_k = 1, \dots, N$, starting from the central core. In this configuration, there are    $(6 - f)$ available connections for the central monomer, 5 connections for each of the    $f$ endpoints (where    $n_k = N$), and 4 connections for the remaining    $N_{\text{total}} - 1 - f$ inner monomers (where    $n_k = 1, \dots, N - 1$). This leads to the following result:
\begin{equation}
\bar{n}_{s,f>1}^{{\rm stiff}}(N_{{\rm total}})=\frac{(6-f)+5f+4f(N-1)}{N_{{\rm total}}} = \bar{n}_{s,f=1}^{{\rm stiff}}(N_{{\rm total}}). \label{stiffstar1}
\end{equation}

The evaluation of the perimeter shell size    $\bar{n}_{a,f>1}^{{\rm stiff}}(N_{{\rm total}})$ {has some nuances}. While the central monomer can adsorb    $6-f$ particles, and each of the end monomers (   $n_k=N$) and inner monomers (   $n_k=2, \dots, N-1$) has coordinatiopn accessibility of 5 and 4, respectively, some particles are shared by monomers with    $n_k=1$ (those adjacent to the central monomer), depending on the mutual orientation of the branches. Indeed, each pair of branches positioned at a    $90^{\circ }$ angle leads to a decrease in    $\bar{n}_{a,f>1}^{\mathrm{s}tiff}$ by one compared to    $\bar{n}_{s,f>1}^{\mathrm{s}tiff}$, since in this case, one site is shared by the first two monomers of these branches. In general, we have:
\begin{equation}
\bar{n}_{a,f>1}^{{\rm stiff}}(N_{{\rm total}})=\bar{n}_{s,f>1}^{{\rm stiff}}(N_{{\rm total}})-\frac{C_f}{N_{{\rm total}}} \label{stiffstarna}
\end{equation}
with $C_f$  calculated on the base of combinatorics: $C_2=12/15$, $C_3=8/3$,  $C_4=14/3$, $C_5=8$, $C_6=12$. As a result,  { mean coordination number of perimeter shell} can be evaluated as \begin{equation}
n_{{\rm bond},f>1}^{{\rm stiff}}(N_{{\rm total}})=\frac{\bar{n}_{s,f>1}^{{\rm stiff}}(N_{{\rm total}})}{\bar{n}_{a,f>1}^{{\rm stiff}}(N_{{\rm total}})}=\frac{2+4N_{{\rm total}}}{2+4N_{{\rm total}}-C_f}.\label{starbond}
\end{equation}

\begin{figure}[b!]
\begin{center}
\includegraphics[width=8cm]{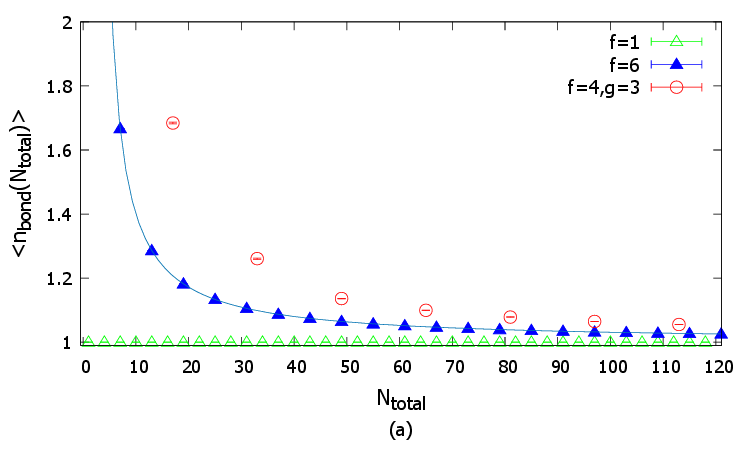}
\includegraphics[width=8cm]{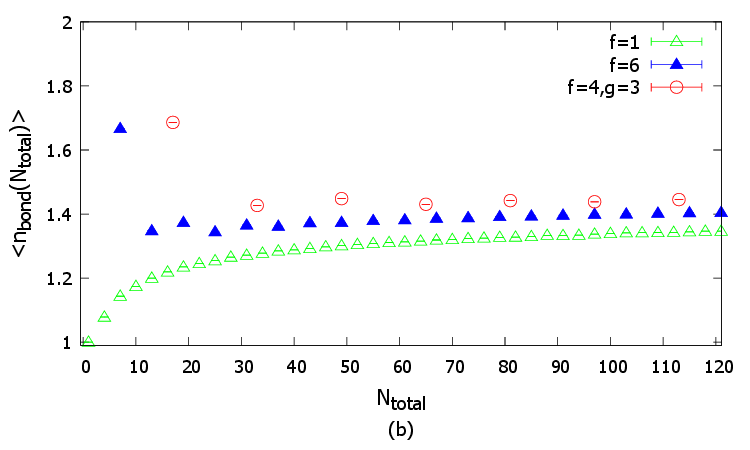}
\includegraphics[width=8cm]{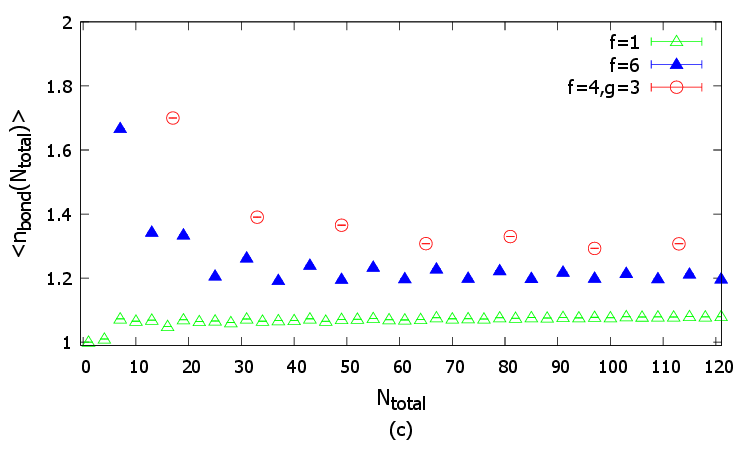}
\end{center}
\caption{\label{nbondn_all}  { Mean coordination number of perimeter shell}, $\langle n_{{\rm bond}} \rangle$, for three representative polymer structures comprising rigid rod chains (a), flexible (b), and tightly packed (c) regimes. Solid line in (a) represent analytical evaluation as given by Eq. (\ref{starbond}).  }
\end{figure}

Our simulation data, presented in Figs. \ref{nan_all}a, \ref{nbondn_all}a, {agree with these theoretical results well}, thereby validating  our simulation approach. 

Based on {these theoretical and simulation results}, we can draw the following conclusions. In such extended, rigid polymer structures,  branching leads to a slight decrease in perimeter shell size $\bar{ n}_{a,f>1,g>1}^{{\rm stiff}}<\bar{n}_{a,f>1,g=1}^{{\rm stiff}}<\bar{n}_{a,f=1}^{{\rm stiff}}$
and an increase of  { mean coordination number of perimeter shell} ${n}_{{\rm bond},f>1,g>1}^{{\rm stiff}}>{n}_{{\rm bond},f>1,g=1}^{{\rm stiff}}>{n}_{{\rm bond},f=1}^{{\rm stiff}}$. 
These effects for rigid structures can be detected only at relatively small values of $N$ and $N_{{\rm total}}$ (for $N_{{\rm total}}\to\infty$ we have $\bar{n}_{s,f,g}^{{\rm stiff}}=\bar{n}_{a,f,g}^{{\rm stiff}}= 4$, $\bar{n}_{{\rm bond},f,g}^{{\rm stiff}}= 1$ independent of $f$ and $g$). 

\begin{figure}[t!]
 \begin{center}
\includegraphics[width=8cm]{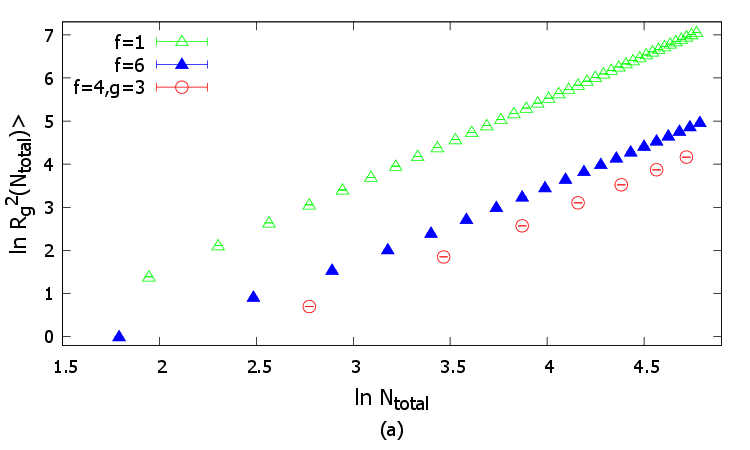}\\
\includegraphics[width=8cm]{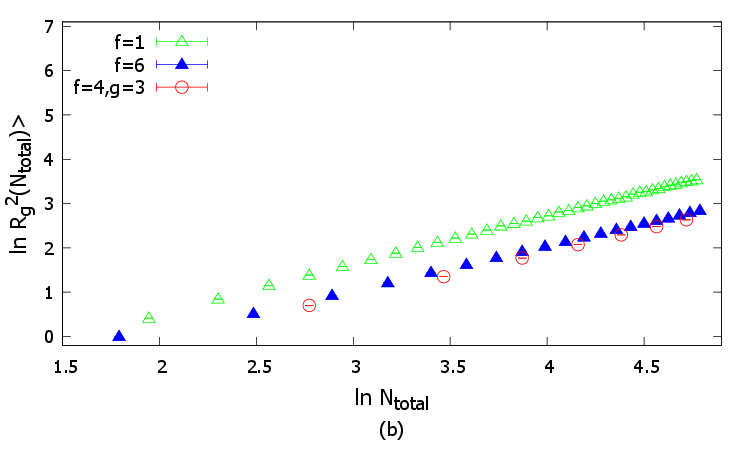}\\
\includegraphics[width=8cm]{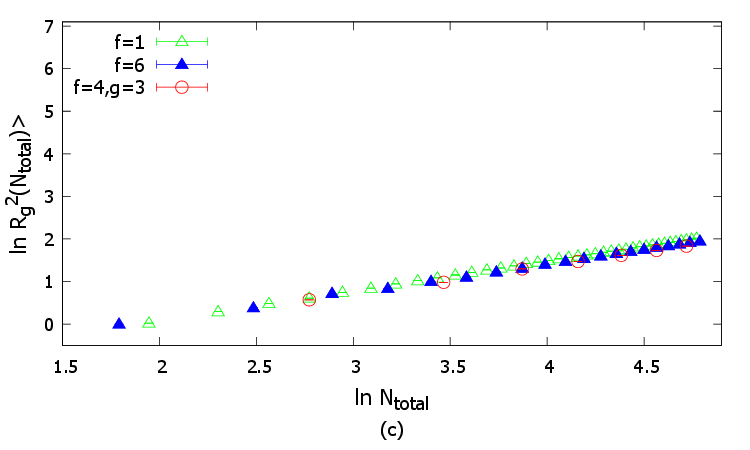}
\end{center}
\caption {\label{rg_all}  { The averaged gyration radius, $\langle R_g^2 \rangle$, for three representative polymer structures comprising rigid rod chains (a), flexible (b), and tightly packed (c) regimes  as function of $N_{{\rm total}}$ in double logarithmic scale.}
}
\end{figure}

Note that {hereafter}, in the majority of figures, we present results for {only} three representative architectures: the linear chain ($  f=1  $), the star polymer with {the maximum possible} functionality ($  f=6  $), and the snowflake polymer with $  f=4  $ and $  g=3  $ twigs. This choice is motivated by the fact that {the intermediate architectures, e.g. star polymers with $f=3-4$, have rather subtle effect on the adsorption-related observables}. 

{  We calculated also the gyration radii for the branched structures in rigid regime, presented in Fig. \ref{rg_all}a.  As expacted, radius of gyration  scales linearly with  $N_{{\rm total}}$ independently of the polymer topology, whereas the value of $R_g^2$ decreases with incresing branching of polymer structure at fixed $N_{{\rm total}}$.}
Additionally, we calculated the averaged asphericity, as defined by (\ref{asfer2}), of a star and snowflake structures in the rigid regime. {While a stiff linear chain, $f=1$, is characterized by the maximum possible asphericity, $A=1$, a star with $f>1$ branches shows gradual decrease of its asphericity with the increase of $f$ (Fig. \ref{asferrod}). The case of $f=6$ is characterized by a perfect global spherical symmetry because all arms run straight along one of the three spatial axes of a $3D$ lattice.}

\begin{figure}[t!]
\begin{center}
\includegraphics[width=7.5cm]{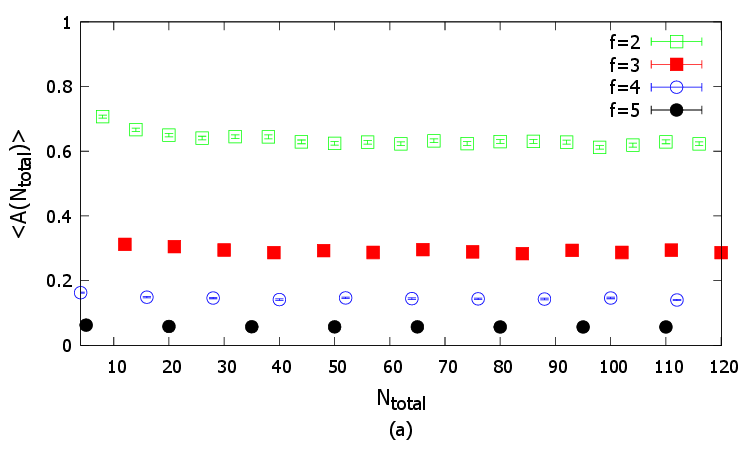}
\includegraphics[width=7.5cm]{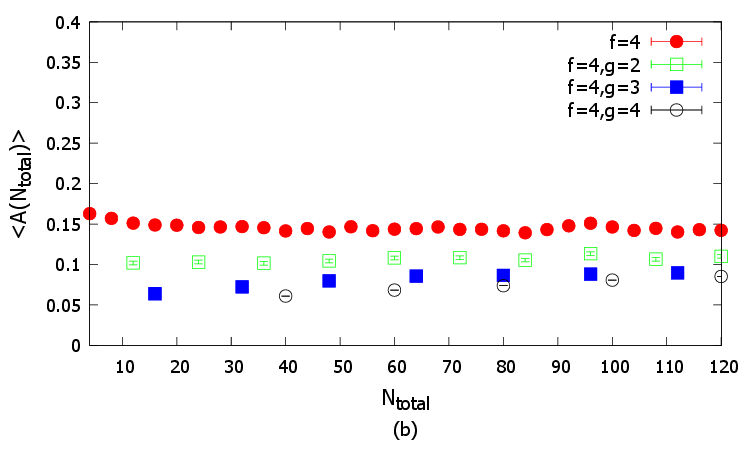}
\end{center}
\caption{\label{asferrod} Averaged asphericity $\langle A \rangle$ for a set of star-like (a) and snoflake-like (b)  structures in rigid regime. {Panel (a) omits the trivial limiting cases    $f=1$ (with    $A=1$) and    $f=6$ (with    $A=0$).} 
}

\end{figure}

\vspace{0.8em}
\noindent{\bf B. Tightly packed state of polymers of various architectures}.\vspace{0.5em}\\
Real-life polymers are often found in collapsed, maximally compact, state, e.g. in the case of a flexible polymer in a poor solvent. In this case, collapse is driven by  strong, solvent-mediated, attraction between monomers.
Typical core of the micellar structure \cite{Negut2023, Abbot2025}, or collapsed polymer brush \cite{Lokesh2024} is exactly in this state. 
{  To mimic collapse of real-life polymers  in our simulation scenario, the compact conformational state is 
generated using specific algorithmic growth rules rather than a continuous variation of physical parameters within a single thermodynamic Hamiltonian.}
This regime also allows for a simplified analytical treatment. { The equilibrium shape of a tightly packed polymer ``globule'' is dictated by the minimization of its surface area, resulting in a shape with the highest possible monomer-monomer contacts.} In continuous $3D$ space, this shape is a sphere. However, in a lattice polymer model, the monomer arrangement is restricted to a grid of discrete positions determined by the lattice symmetry, which affects the globule's shape.
The analog of a sphere of radius $r$ on a discrete cubic lattice  is most naturally considered to be a cube with an edge length  $n_{\text{edge}}=2r$, as the simplest irreducible closed volume for the Chebyshev metric \cite{serra}. Indeed, it is defined by the condition for the maximum coordinates: $\max(|x|, |y|, |z|) \le r$. For the case of a polymer composed of $N_{{\rm total}}$ monomers we should take into  account, that perfect cube can be formed only if the edge length as given by $n_{\text{edge}}=N_{{\rm total}}^{1/3}$ is integer, otherwise the compact rectangular cuboid, octahedron or rhombododecahedron will be formed.  The perfect cubic ground state for lattice polymers in 3D are observed e.g. in Ref. \cite{Vogel2007} at very low temperature limit (extremely poor solvent). The temperature fluctuations and entropic effects lead to ``smearing'' of this  highly symmetric crystalline structure.


\begin{figure}[b!]
\begin{center}
\includegraphics[width=7.5cm]{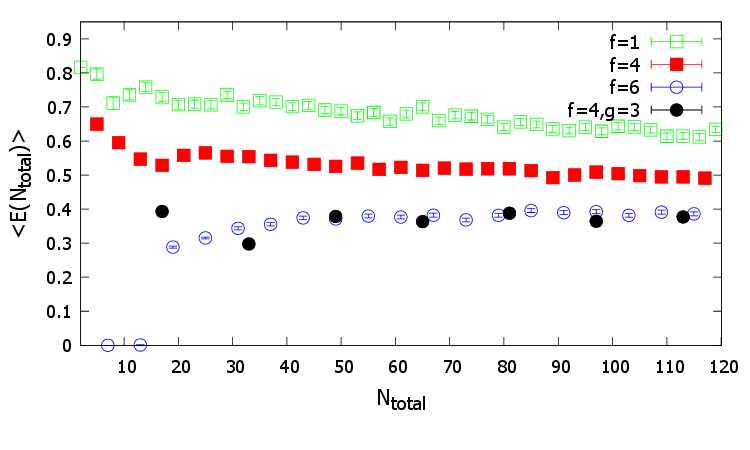}
\end{center}
\caption{\label{pora} Averaged bounding-box anisotropy  index $\langle E \rangle$   for representative polymer structures in tighlty packed  regime. }
\end{figure}

{ To establish a basic theoretical background for analyzing the perimeter properties in the maximally compact state, we consider a simple limiting geometric reference: a perfectly filled, non-fluctuating cubic cluster composed of \(N_{{\rm total}}\) monomers. It is important to emphasize that this cuboidal configuration does not represent the true thermodynamic equilibrium ground state of a collapsed branched polymer structure  under nearest-neighbor attraction, which is highly non-trivial. Instead, it serves purely as a rigid geometric upper bound for a completely packed discrete volume.
Under this strict limiting assumption of a perfect cube with an integer edge length \(n_{edge} = N_{{\rm total}}^{1/3}\), the distribution of monomers is as follows:} there are 8 corner particles with 3 available connections each,    $12(n_{\text{edge}}-2)$ particles along the edges with 2 connections,    $6(n_{\text{edge}}-2)^2$ particles on the faces with only 1 open connection, and    $(n_{\text{edge}}-2)^3$ interior particles that are completely 'screened'. Inside the globule, the polymer is packed without internal voids, representing a 'frozen' or 'globular' phase. As a result:
\begin{equation}
\bar{n}_s^{{\rm globule}}=\frac{8\cdot3+12(n_{{\rm edge}}-2)\cdot2+6(n_{{\rm edge}}-2)^2}{N_{{\rm total}}}=\frac{6}{N_{{\rm total}}^{1/3}}. \label{nsglob}
\end{equation}
This value slowly approaches $0$ in the limit of infinite $N$. Straightforwardly we obtain   $\bar{n}_a^{{\rm globule}}=\bar{n}_s^{{\rm globule}}$, and correspondingly $n_{{\rm bond}}^{{\rm globule}}=1$.  One should note that this idealized cuboidal configuration serves purely as a basis for comparison with our numerical data. 

\begin{figure}[t!]
	\begin{center}
		\includegraphics[width=7.5cm]{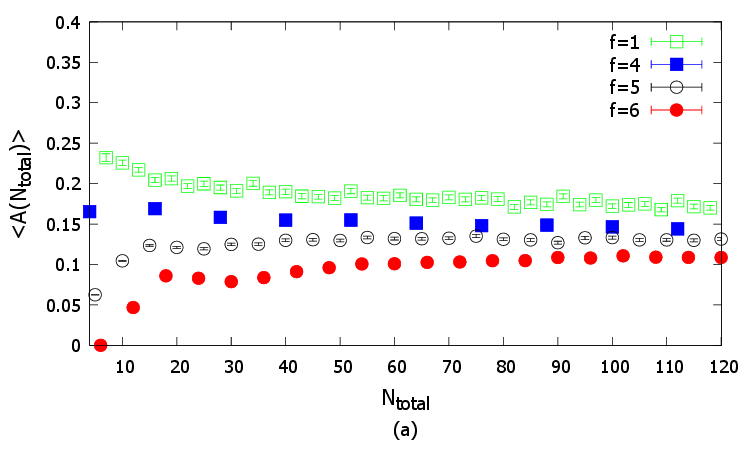}
		\includegraphics[width=7.5cm]{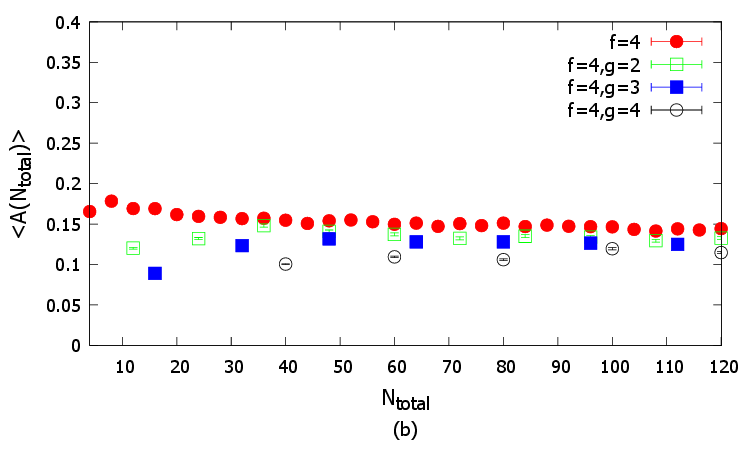}
	\end{center}
	\caption{\label{asferglob} Averaged asphericity $\langle A \rangle$ for a set of star-like (a) and snowflake-like (b)  structures in tightly packed  regime. }
\end{figure}

Now we proceed to the simulation results. 
{  The gyration radii for the branched structures under consideratin are presented in Fig. \ref{rg_all}c.  The fitting of resulting curves restores the value of scaling exponent $\nu=0.334(1)$ known for the compact globular polymer state \cite{desCloiseaux1991}, independently of the topology of considered polymer structures.}
Let us begin with some general considerations about the shape of the cluster formed by polymer structures in the compact regime. The linear dimensions of the cluster (``bounding box'') denoted as    $size_{x}$,    $size_{y}$, and    $size_{z}$, define the minimum orthogonal volume that completely encompasses the set of $N_{{\rm total}}$ monomers within the three-dimensional cubic lattice. Due to the discrete nature of the grid, these parameters represent the total number of lattice sites along each coordinate axis and are computed using the extreme coordinates of the set of monomers of  polymer structure in collapsed states. To quantitatively assess the spatial anisotropy and geometrical distortion of the shape of collapsed polymer structure, we introduce an integrated scalar metric designated as the { bounding-box anisotropy index} ($E$):
\begin{equation}
E = \sqrt{\frac{(r_1 - 1)^2 + (r_2 - 1)^2 + (r_3 - 1)^2}{3}}, \label{eindex}
\end{equation}
with the {axial} aspect ratios    $r_1 = size_x/size_y$,    $r_2 = size_x/size_z$, and    $r_3 = size_y/size_z$. This non-negative metric provides a rigorous, coordinate-independent measure of shape variation; a value of    $E = 0$ corresponds to a cubic shape, whereas progressively higher values indicate an increasing degree of structural elongation towards a cuboidal shape.
{  Note that parameter (\ref{eindex}) serves primarily as a convenient operational tool to quantify how the bounding volume of the collapsed cluster deviates from a perfect cube (the natural irreducible volume in Chebyshev metric on a cubic grid) toward a cuboidal shape.}
Our simulation data for the averaged value of this parameter, as given in Fig. \ref{pora}, demonstrate a deviation of the shape of compact cluster from the perfect cube, more pronounced for the linear polymer structures. 
This tendency is further supported by evaluating the asphericity, $A$, of the {corresponding polymer structures} (Fig. \ref{asferglob}). In its collapsed state, a linear polymer chain with    $f=1$ forms the most ``unspherical'' (or rather, within the framework of a regular lattice, the most ``uncubical'') cluster. Increasing the  branching  leads to enhanced shape symmetry in the resulting cluster.  {  Thus both independent metrics, based on based on bounding-box anisotropy index $E$ and rotationally invariant asphericity $A$, are in agreement: the linear chain (\(f=1\)) forms the most elongated/least spherical structure in the tighly packed state, whereas   branching drives the system toward globally symmetric shape with  asphericity value close to zero.}

{The internal structure and effective local cavity structure of the collapsed polymer adsorbent strongly influence its adsorption efficiency. While a full characterization of local cavity structure involves complex cavity size distributions and interconnectivity, we can estimate local site availability by analyzing the distribution of the mean coordination number   $n_{\text{bond}}$. The perimeter shell sites can be occupied by adsorbed particles, making their distribution (shown in Fig. \ref{probcollapse}a) a direct measure of structural accessibility. Note that only data for    $n_{\text{bond}}\geq 3$ are shown here, to concentrate specifically on the probability of forming local geometric cavities— specifically, unoccupied lattice sites surrounded by 4 or 5  monomers.  
This agrees with earlier simulations \cite{Sikorski1998}, which demonstrated that even in the collapsed state, each arm of a star polymer remains relatively extended due to the high monomer density near the central branching point. 
{   We should mention that our model evaluates the static local geometric accessibility based on frozen snapshots of the polymer matrix, which does not account for the kinetics of the adsorption process.  In real polymer systems, the size of spatial cavities within collapsed or branched architectures will depend  on the competition between the timescales of adsorbate diffusion and polymer chain fluctuations.}

\begin{figure}[t!]
\begin{center}
\includegraphics[width=7.5cm]{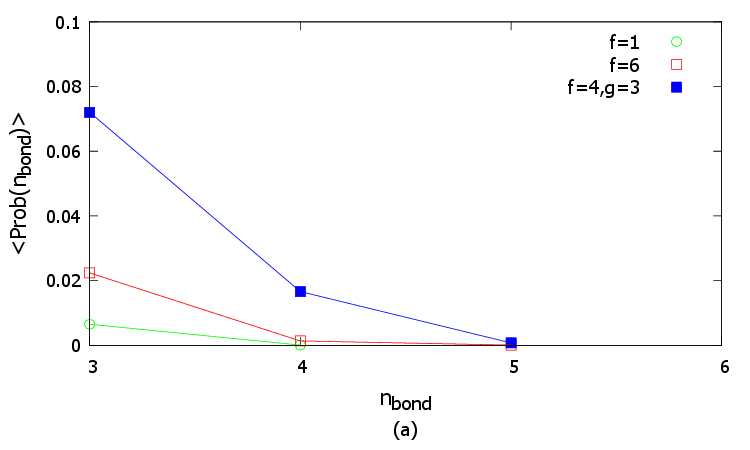}
\includegraphics[width=7.5cm]{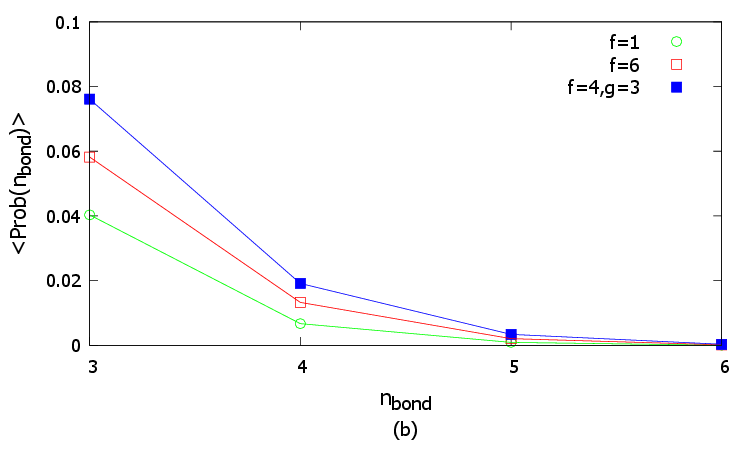}
\end{center}
{\caption{\label{probcollapse}Probability distribution of the number of contacts,    $n_{{\rm bond}}$ with $n_{{\rm bond}}\ \geq 3$  between adsorbed particles and monomers, for various polymer structures in the tightly packed state (a) and flexible  state (b) at $N_{{\rm total}}=49$.
}}
\end{figure}

The general tendency is preserved for the geometry characteristics of polymer structures in tightly packed state that are related to their adsorbing abilities , observed earlier for their counterparts comprising rod-like chains:  branching leads to a slight decrease of perimeter shell size: $\bar{n}_{a,f>1,g>1}^{{\rm globule}}<\bar{n}_{a,f>1,g=1}^{{\rm globule}}<\bar{n}_{a,f=1}^{{\rm globule}}$
and an increase of  { mean coordination number of perimeter shell} $n_{{\rm bond},f>1,g>1}^{{\rm globule}}>n_{{\rm bond},f>1,g=1}^{{\rm globule}}>n_{{\rm bond},f=1}^{{\rm globule}}$ (cf. also our simulation data presented in Figs. \ref{nan_all}c, \ref{nbondn_all}c). 

\vspace{0.8em}
\noindent{\bf C. Flexible coils of various architecture}\vspace{0.5em}\\
{Typical real-life  polymers are often found in a swollen state (flexible polymer in a good solvent, etc.). Our simulation results for this case} are presented in Figs. \ref{nan_all}b and \ref{nbondn_all}b. Regarding the impact of branching on the observables, the main tendencies remain consistent with those observed previously: the maximum perimeter shell size    $\bar{n}_{a}^{\text{coil}}$ decreases with increasing branching, while the  { mean coordination number of perimeter shell}   $\langle n_{\text{bond}}^{\text{coil}} \rangle$ conversely increases. { These tendencies can be clearly followed in Fig. \ref{nan_flex}, which summarizes our data obtained for a set of star-like and snowflake-like structures at fixed total molecular weight $N_{{\rm total}}=121$ for comparison.  }

\begin{figure}[t!]
 \begin{center}
\includegraphics[width=8cm]{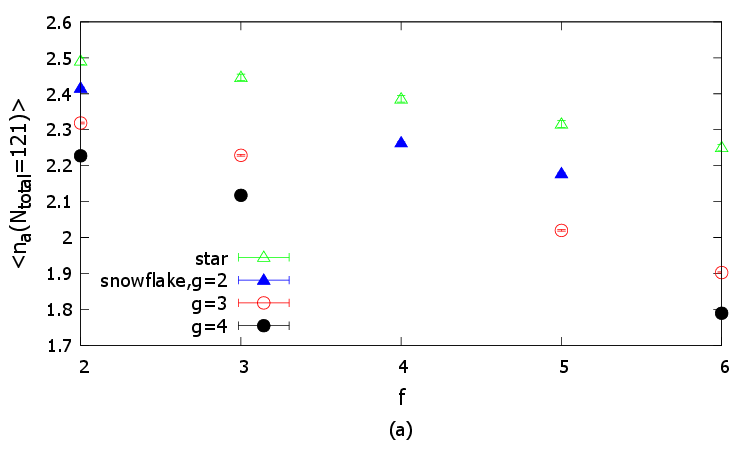}\\
\includegraphics[width=8cm]{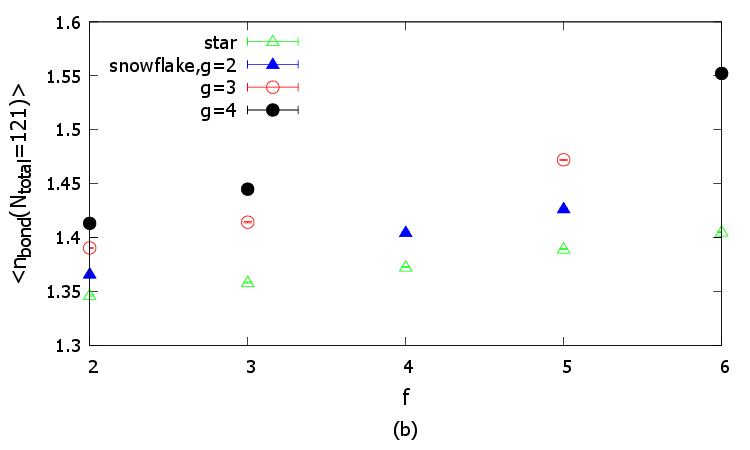}
\end{center}
\caption {\label{nan_flex}  Perimeter shell size $\langle \bar{n}_{a} \rangle$ (a) and mean coordination number of perimeter shell  $\langle n_{\text{bond}}\rangle$ (b) for a set of star and snowflake structures at fixed $N_{{\rm total}}=121 $ in  flexible regime as function of $f$.
}
\end{figure}

{  In this study we do not evaluate local cavity structure of polymers explicitly, via distribution of pore sizes, their interconnectivity, etc. Instead we consider related characteristic such as} the distribution function of the number of bonds between unoccupied lattice sites  and monomers,    $n_{\text{bond}}$, as presented in Fig. \ref{probcollapse}b. The general tendency {is inline with  the case of tightly packed regime} (cf. Fig. \ref{probcollapse}a: an increased degree of branching supports the higher probability of forming cavities, unoccupied lattice sites surrounded by more than 4 (up to 6) monomers; however this effect is less pronounced comparing to the tighly packed state.

\begin{figure}[h!]
\begin{center}
\includegraphics[width=7.5cm]{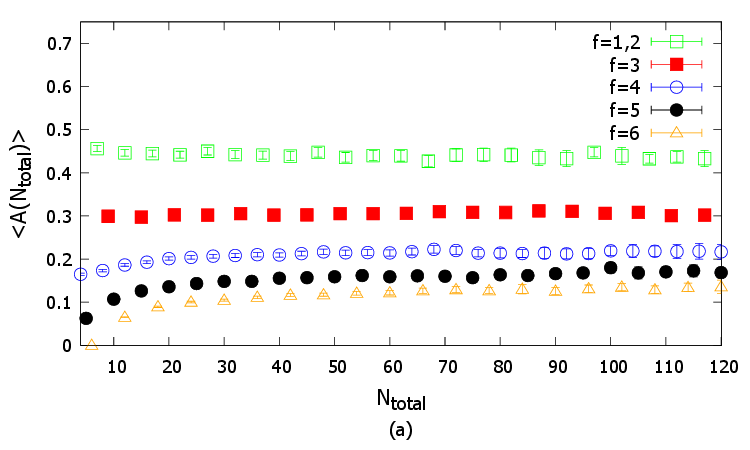}
\includegraphics[width=7.5cm]{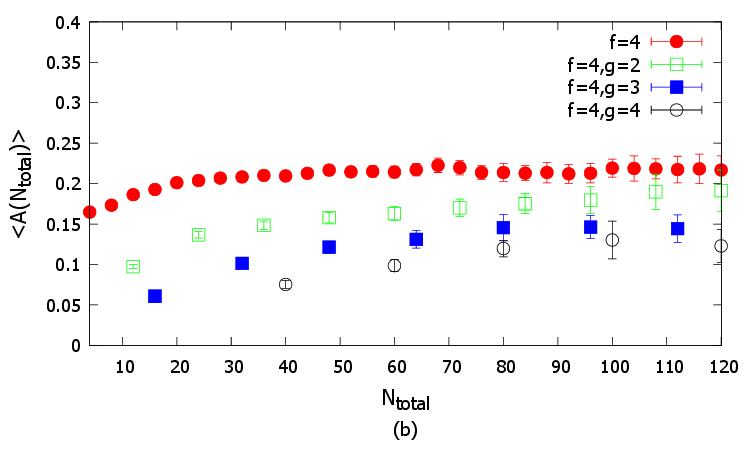}
\end{center}
\caption{\label{asferflex} Averaged asphericity $\langle A \rangle$ for a set of star-like (a) and snoflake-like (b)  structures  in  flexible  regime. }
\end{figure}

For completeness, we also analyzed 
{  the gyration radii for the representative set of polymer topologies, as presented in Fig. \ref{rg_all}b.  The fitting of resulting curves restores the value of scaling exponent $\nu=0.586(1)$ known for the scaling of polymers in good solvent regime, as given e.g. in  \cite{Guida1998}.}
The data for the asphericity parameter    $\langle A \rangle$ of the polymer structures under consideration in the good solvent regime are presented in Fig.~\ref{asferflex}a, they agree well with the corresponding values obtained earlier for star-like polymers with up to    $f=6$ in Ref.~\cite{Bishop1991}. The results for a representative snowflake polymer structure with a fixed    $f=4$ and various    $g$ (Fig.~\ref{asferflex}b) support the general tendency toward an increasing symmetry of the typical polymer configuration with a higher degree of branching.

\begin{figure}[t!]
	\begin{center}
		\includegraphics[width=8cm]{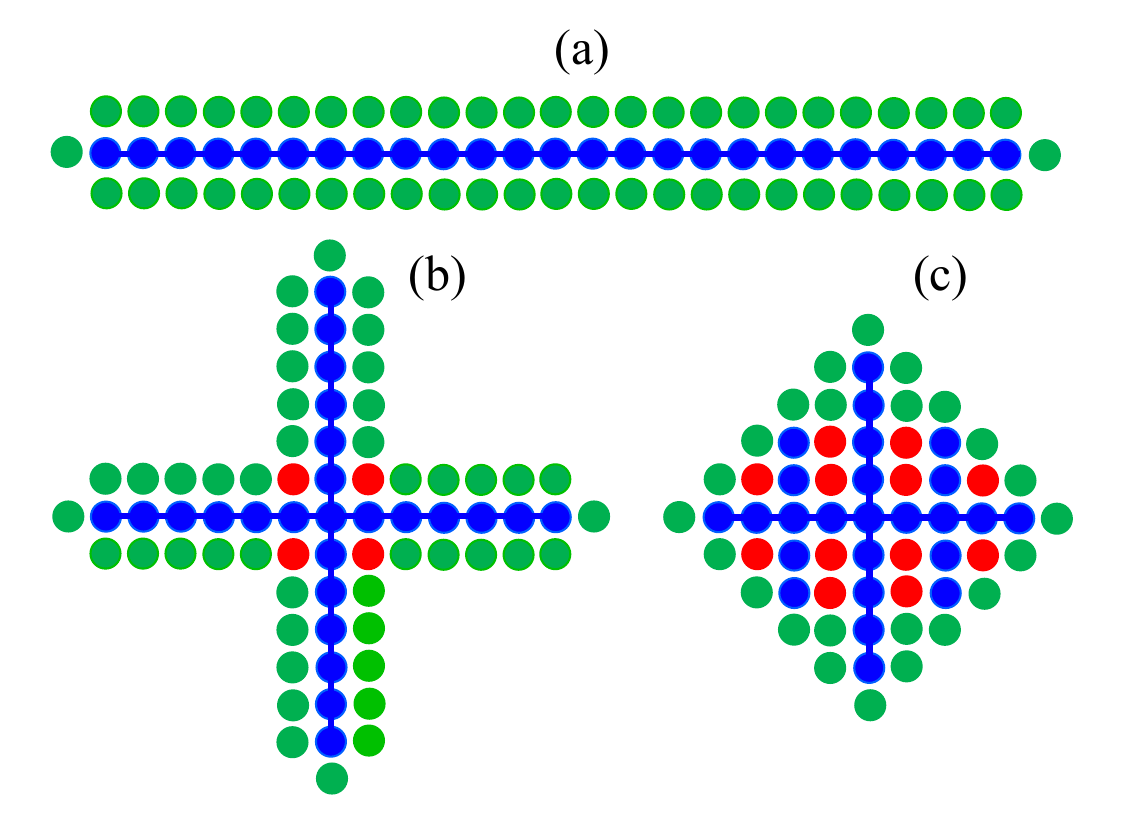}
	\end{center}
	\caption{\label{illustration_rigid}{Two-dimensional ($2D$) visualization of the spatial arrangement of perimeter shell of polymer structure comprising rod-like chains. Panels show: (a) linear polymer, (b) a star-like polymer, and (c) a partial snowflake polymer. Perimeter sites are depicted in blue, perimeter shell sites connected via a single bond in green, and those connected via a double bond in red. The total polymer length is identical in all three cases and equals    $25$ monomers.}}
\end{figure}

\begin{figure}[b!]
	\begin{center}
		\includegraphics[width=6cm]
		{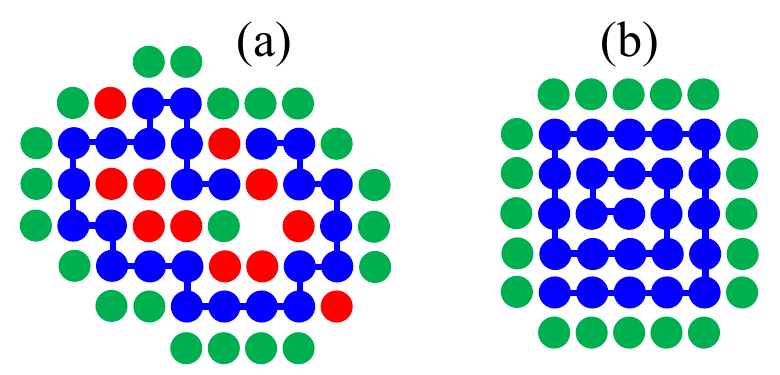}
	\end{center}
	\caption{\label{illustration_SAW_glob}{Visualization of the effect of flexibility for the case of a linear chain with $25$ monomers. (a) a flexible constituent chains, (b) tightly packed state}.}
\end{figure}

\vspace{0.8em}
\noindent{\bf D. Comparison between the adsorbents of various architectures comprising rod-like, flexible, or tighly packed chains}\vspace{0.5em}\\
Let us overview the three regimes discussed above, starting with the data for perimeter shell size presented in Fig. \ref{nan_all}. Polymers comprising rigid-rod chains are the most effective in terms of the geometric accessibility of their monomers for adsorption of particles: as the total molecular weight    $N_{\text{total}}$ increases,    $\langle \bar{n}_a^{\text{rod}} \rangle$ reaches a value of    $4$ in the large-$N$ limit for all branching topologies (Fig. \ref{nan_all}a). In the case of the structures comprising flexible coiled chains (Fig. \ref{nan_all}b), the values of    $\langle \bar{n}_a^{\text{coil}} \rangle$ for each topology are considerably smaller compared to those in the rigid state.  In the case of tightly packed polymeric structures, the perimeter shell size is, as expected, the lowest. At small    $N$, the tightly packed structures (globules) formed by branched structures are more 'porous' and thus characterized by smaller values of    $\langle \bar{n}_a^{\text{globule}} \rangle$ compared to linear chains. With increasing    $N$, the differences between the    $\langle \bar{n}_a^{\text{globule}} \rangle$ values for various topologies practically disappear.

We continue with the results for the  { mean coordination number of perimeter shell} shown in Fig. \ref{nbondn_all}. Structures comprising rod-like chains are clearly the weakest in this regard: as the total molecular weight    $N_{\text{total}}$ increases,    $\langle n_{\text{bond}}^{\text{rod}} \rangle$ reaches a limiting value of    $1$ in the large-$N$ limit for any branching topology (Fig. \ref{nbondn_all}a). The structures in a flexible regime (Fig. \ref{nbondn_all}b) are characterized by the highest values of    $\langle n_{\text{bond}}^{\text{coil}} \rangle$. Note that while the  {mean coordination number of perimeter shell} of a linear flexible chain slightly increases with    $N$, whereas the values for branched structures comprising  flexible chains,  decrease with    $N$. This decrease is due to the diminishing role of the mutual crowding effect of branches in the vicinity of the branching cores.   
In the tightly packed regime, the larger values of    $\langle \bar{n}_{\text{bond}}^{\text{globule}} \rangle$ of branched structures comparing with that of linear one is caused by more pronounced ``local cavity structure'' of clusters formed by them; this difference diminishes with increasing    $N$. The zigzag-like behavior of the number of available open sites (pronounced for     $n_{\text{bond}}$ as a function of arm length    $N$, as shown in Figs. \ref{nbondn_all}), is primarily due to the discrete, parity-dependent nature of the lattice, combined with the excluded volume effect at the crowded central node. This occurs because the number of available adjacent lattice sites depends heavily on whether    $N$ is odd or even, reflecting the bipartite nature of the cubic lattice; monomers on even steps often face different crowding constraints than those on odd steps \cite{Campbell2021,Wilkinson1986}. For large    $N$, this zigzag effect is gradually vanishing.

{ As one can see from  Figs.~\ref{nan_all} and \ref{nbondn_all}, the density of branching in polymer structures strongly affects its geometric properties related to adsorption. To analyze this effect quantitatively, we introduce a  dimensionless branch-point density that takes into account both the number of branching centers and  their functionalities, according to 
\begin{equation}
\rho_{f,\,g}=\frac{(f-2)+f(g-1)}{N_{{\rm total}}}.
\end{equation}
 Thus, one can compare structures with various branching topology and fixed $N_{{\rm total}}$. Whereas for the linear chain we have the minimum value
$\rho_{f=1,2,\,g=0}=0$,  it increases
 for  star structures with increasing $f$ (e.g. for $f=6$-branch star
 $\rho_{f=6,\,g=0} = 4/N_{{\rm total}}$), and grows further  for the snowflake  topologies (e.g. at $f=4$, $g=3$ one has $\rho_{f=4,\,g=3} = 10/N_{{\rm total}}$.  The systematic increase of $\rho _{f,\,g}$ enhances local monomer crowding, and thus reduces the perimeter shell size  $\langle \bar{n}_{a} \rangle$, but increases the   mean coordination number of perimeter shell, $\langle n_{{\rm bond}} \rangle$.} The effect is very prominent for the case of systems comprising rod-like chains, frames (a) in both Figs.~\ref{nan_all} and \ref{nbondn_all}, and can be more easily visualized for this case. Mutual arrangement of a polymer and of all possible impurities adsorbed on it, is shown in Fig.~\ref{illustration_rigid} for the $2D$ case and the same number of monomers in all three cases equal to $25$. Polymer sites are shown in blue, adsorbed impurities connected via a single bond in green, and adsorbed impurities connected via a double bond in red. It is clear, that in the vicinity of each branching point there are sites, available for adsorption, that are neighbors to more than one chain. As the result, the total number of adsorbed impurities for branched architectures decreases and is equal to $52$, $48$, and $36$ in frames (a), (b), and (c) of Fig.~\ref{illustration_rigid}, respectively. However, the impurities located at these sites (shown in red) form more than one link with the polymer, and their fraction for branched architectures grows and is: $0$, $0.083$, and $0.333$ for the same respective cases. This {explains the geometric origins for} the effect of branching on the adsorption characteristics as seen in frames (a) of Figs.~\ref{nan_all} and \ref{nbondn_all}. 

The effect of polymer flexibility is illustrated in Fig.~\ref{illustration_SAW_glob} for the linear polymer of $25$ monomers. One can see that the self-avoidance of a flexible chain, see frame (a), has the similar effect on both $\langle \bar{n}_{a} \rangle$ and $\langle n_{{\rm bond}} \rangle$ as chemical branching. Indeed, bending of a chain promotes formation of internal geometric cavity, where the restricted number of impurities can be adsorbed, albeit with their strong bonding with two or three monomers. For the case of a tightly packed state, see frame (b), we depicted the case of an ideal $2D$ packing with no voids in the interior of a chain. The situation is different in $3D$, but the general trend for both $\langle \bar{n}_{a} \rangle$ and $\langle n_{{\rm bond}} \rangle$ to stay on a low side, as is seen Figs.~\ref{nan_all} and \ref{nbondn_all}, is the same.

{
The experimental adsorption capacity is affected by both availability of adsorbing sites (characterized by    $\langle \bar{n}_a \rangle$), and by the  { mean coordination number of perimeter shell},    $\langle n_{\text{bond}} \rangle$. The importance of the latter can be envisaged in the case, when the minimum  { mean coordination number of perimeter shell}, required for stable binding of impurities over prolonged time, can not be provided by a single monomer-impurity bond, but requires double or triple bonding. We note, that despite the similarities between the effect of branching and coiling of a flexible linear chain on    $\langle n_{\text{bond}} \rangle$, the effect from the latter is always weaker, see Fig.~\ref{nbondn_all}. Branched polymer is characterized by a fixed set of branching points resulting in formation of geometric cavities, adjacent to each of them. These allow for double- or triple-bonds for impurities, thus increasing the average value of    $\langle n_{\text{bond}} \rangle$. The effect is subtle for the set of regular architectures, each containing just a few branching points, discussed here. But it will be amplified enormously for the case of a hyperbranched polymer, with a huge number of branching points. In the case of a long linear chain, formation of such number of cavities requires very specific conformations that are improbable due to entropic considerations. This interpretation may serve as an explanation of the experimentally observed higher adsorption capacity of branched polymers compared with the linear ones, both in chelation \cite{Kobayashi1987, Yi2024}, and flocculation \cite{Blanco2009}.
}

\section{Conclusions and Discussions}\label{IV}

We studied the perimeter shell properties for branched macromolecules (star and snowflake polymers) that can be associated with their adsorbing abilities. The study involves both theoretical calculations and  Monte Carlo simulations  on a cubic lattice. We compared { three representative classes of polymer conformations: polymers comprising rod-like chains, flexible coil states, and tightly packed states related to the collapsed globular regime in poor solvent}. The quantitative results for a set of adsorption characteristics, such as perimeter shell size per monomer $\bar{n}_a$  and  { mean coordination number of perimeter shell} $n_{{\rm bond}}$ are obtained. 
These lattice geometry properties allow us to connect the microscopic properties to real-world experimental observables. 
{  Indeed, the perimeter shell size defines the theoretical maximum adsorption capacity  in the limit of non-interacting adsorbate: the actual amount of captured particles is expected to scale linearly with their low concentration $c$, in accordance with the low-density phenomenological assumption, limited to low concentration $c$ and supported by some previous simulations. }
This mapping allows us to interpret experimental data for disordered or defect-rich polymer matrices and helps in designing responsive materials.

Our results quantitatively demonstrate that both polymer topology (degree of branching) and conformational state of their constituent chains are important in determining the geometric accessibility of its monomers that is related to 
the overall adsorption efficiency of these macromolecules.
Across all regimes, higher branching (from linear chains to stars and snowflake architectures) leads to a moderate decrease in the perimeter shell size $\bar{n}_a$ while simultaneously enhancing the  {mean coordination number of perimeter shell} $n_{{\rm bond}}$. This is caused by  the higher local monomer density near branching cores, leading to formation of local geometric cavities and shared adsorption sites between neighboring branches. In the case of polymers comprising rigid chains, the highest perimeter shell size is observed, approaching the theoretical limit of $4$ available sites per monomer for large $N$ in $3D$ with excellent agreement between simulations and analytical predictions. Polymer structures comprising flexible chains show intermediate perimeter shell size but the strongest particle bonding due to their crumpled conformations that facilitate multi-monomer contacts. However, branched structures form more porous globules (which we observe indirectly, via a set of evaluated characterisitcs), making their internal sites more accessible than those of linear chains.
Our simulations confirm that this internal local cavity structure enhances site accessibility.  

We also calculated the average asphericity parameter $\langle A \rangle$ across all three regimes. In the rigid regime, the linear chain is fully anisotropic ($A=1$). Increasing the core functionality    $f$ makes the molecule more symmetric, until the structure is characterized by zero asphericity ($A=0$) at    $f=6$. A similar trend holds for the tightly packed compact state, where branched architectures form more isotropic and compact clusters than linear chains. In the flexible regime, branching likewise promotes sphericity, which agrees with earlier theoretical and simulation studies.
These shape characteristics explain  the molecular mechanisms behind the observed adsorption behavior: more symmetric conformations enhance the multi-contact binding but slightly reduce the total accessible surface area.

It can be worth questioning whether  our main conclusions and findings, observed in the frameworks of  lattice model, are applicable to  off-lattice (continuous space) polymer representations. Indeed, a simple cubic lattice model presents serious  limitations, such as discrete coordination numbers and strict directional bonding. However our main conclusions  rely on universal topological features. In an off-lattice continuous model, the central physical mechanisms, such as crowding of monomers around the branching cores and 
steric competition between branches, are topologically driven. Thus, though we expect the quantitative values of observables (such $\hat{n}_a$ and $n_{{\rm bond}}$) to differ from their counterparts for the case of continuous models, the qualitative tendencies of the impact of branched architecture and the role of chain flexibility will remain invariant.


Our forthcoming studies will extend this analysis to higher-generation dendrimers and explore the dynamics of the adsorption process.

\section*{Acknowledgments}
The work was supported by the National
Research Foundation of Ukraine Project ``Functional efficiency of complex multibranched polymers in adsorption-based applications: computer simulations and predictions'' No. 2025.07/0366.

\bibliography{Papers}
\end{document}